\documentclass[fleqn,12pt,twoside]{article}
\usepackage{espcrc1}

\usepackage{graphicx}
\usepackage[figuresright]{rotating}

\newcommand{\AmS}{{\protect\the\textfont2
  A\kern-.1667em\lower.5ex\hbox{M}\kern-.125emS}}

\title{Matter-wave interference, Josephson oscillation and its disruption
in a Bose-Einstein
condensate on an optical lattice}

\author{Sadhan K. Adhikari\thanks{Supported in part by
the CNPq of Brazil} 
\\
\vskip .2cm
Instituto de F\'isica Te\'orica,
Universidade Estadual paulista,\\ 01405-900 S\~ao Paulo, S\~ao Paulo,
Brazil }
       
\begin{document}

\maketitle

\begin{abstract} Using the axially-symmetric time-dependent mean-field 
Gross-Pitaevskii equation we study the Josephson oscillation in a
repulsive Bose-Einstein condensate trapped by a harmonic plus an
one-dimensional optical-lattice potential to describe the experiments by
Cataliotti {\it et. al.} [Science 293 (2001) 843, New J. Phys. 5 (2003)
71.1]. After a study of the formation of matter-wave interference upon
releasing the condensate from the optical trap, we directly investigate
the alternating atomic superfluid Josephson current upon displacing the
harmonic trap along the optical axis.   
The Josephson current is found to be
disrupted upon displacing the harmonic trap through a distance greater
than
a critical distance signaling a superfluid to a classical insulator
transition in the condensate.  \end{abstract}

\section{INTRODUCTION}
The experimental loading of a cigar-shaped Bose-Einstein condenate
(BEC) in both one-  \cite{1,2,3,4}  and three-dimensional  \cite{5} 
periodic optical-lattice 
potentials  
has allowed to study the quantum phase
effects on a macroscopic scale such as interference of matter
waves.  There have been several theoretical studies on 
a BEC in a one-  \cite{6,7,8} and three-dimensional
\cite{9}
optical-lattice
potentials.
The phase coherence between
different sites of a trapped BEC on an optical lattice has been
established in  recent experiments
\cite{1,2,3,4,5} through the formation of
distinct interference pattern when the traps are removed.

In the experiment on matter-wave interference
from two pieces of coherent BEC, an
interference pattern comprised of a large number of dark and bright
patches   is formed \cite{10}. This is similar to the well-known
double-slit interference   pattern in optics. As the number of
slits in
the experiment on interference of light is increased, the number of bright
patches in the interference pattern is reduced and one has only a few
prominent bright patches for interference of light from an optical
grating with very large number of narrow slits.
A similar
phenomenon has emerged in  matter-wave interference
from a BEC in an optical-lattice trap created by a standing-wave laser
field, which can be considered to be a large
number of  coherent sources of matter wave. 
When a BEC trapped in an axially-symmetric
harmonic and an one-dimensional optical-lattice trap  is released from the
joint traps a definite interference pattern composed of three peaks is
usually formed.   With the
increase of lattice spacing the interference pattern evolves to
$(2{\cal N}+1)$
peaks with ${\cal N}$ peaks symmetrically located in a straight
line
on opposite sides
of a central peak. 

 Cataliotti {\it et al.}  \cite{2}  prepared a BEC on a joint harmonic
plus an optical-lattice
trap. 
Upon displacing the harmonic trap along the optical lattice, the BEC
was found to execute the Josephson oscillation by quantum tunneling  
through  the optical-lattice barriers.
In a later experiment they \cite{3} found that
for a larger displacement of the harmonic trap the Josephson oscillation
is disrupted. 
Their  measurement of the Josephson
oscillation  was rather indirect and based on the existence of phase
coherence and the formation of interference pattern upon release from the
traps. 
They adopted the indirect  procedure as the expansion of the BEC upon
release from the
traps facilitates the observation. However, this is not quite necessary in
numerical simulation, where one can directly identify  the Josephson
oscillation and its disruption without resorting to an expansion. A
preliminary theoretical study in this topic was based on an expansion of
the BEC as in the experiment \cite{7}. Here, we investigate the 
Josephson oscillation and its disruption based  on a direct study of the
BEC. We compare  these results with experiments \cite{2,3} as well as with
previous 
results \cite{7} based on an expansion of the BEC upon release from the
traps.

\section{MEAN-FIELD MODEL AND RESULTS}
The time-dependent BEC wave
function $\Psi({\bf r};\tau)$ at position ${\bf r}$ and time $\tau $
is described by the following  mean-field nonlinear Gross-Pitaevskii (GP)
equation
\cite{7,8}
\begin{eqnarray}\label{a} \left[- i\hbar\frac{\partial
}{\partial \tau}
-\frac{\hbar^2\nabla^2   }{2m}
+ V({\bf r})
+ {\cal G}N|\Psi({\bf
r};\tau)|^2
 \right]\Psi({\bf r};\tau)=0,
\end{eqnarray}
where $m$
is
the mass and  $N$ the number of atoms,
 $ {\cal G}=4\pi \hbar^2 a/m $ the strength of interaction,
 with
$a$ the scattering length.  In the presence of the combined
traps      $  V({\bf
r}) =\frac{1}{2}m \omega ^2(\rho ^2+\nu^2 y^2) +V_{\mbox{opt}}$ where
 $\omega$ is the angular frequency of the harmonic trap
in the radial direction $\rho$,
$\nu \omega$ that in  the
axial direction $y$,  and $V_{\mbox{opt}}$ is
the optical-lattice potential.
The axially-symmetric wave function
can be written as
$\Psi({\bf r}, \tau)= \psi(\rho,y,\tau)$, where $0\le  \rho < \infty$  and 
$-\infty <y<\infty $.
Transforming to
dimensionless variables $\hat \rho =\sqrt 2 \rho /l$,  $\hat y=\sqrt 2
y/l$, $t=\tau
\omega, $
$l\equiv \sqrt {\hbar/(m\omega)}$,
and
${ \varphi(\hat \rho,\hat y;t)} \equiv   \hat \rho\sqrt{{l^3}/{\sqrt
8}}\psi(\rho,y;\tau),$  Eq. (\ref{a}) becomes \cite{7,8}
\begin{eqnarray}\label{d1}
\biggr[-i\frac{\partial
}{\partial t} -\frac{\partial^2}{\partial
\hat \rho ^2}+\frac{1}{\hat \rho}\frac{\partial}{\partial \hat \rho}
&-&\frac{\partial^2}{\partial
\hat y^2}
+\frac{1}{4}\left(\hat \rho^2+\nu^2 \hat y^2\right)
+\frac{V_{\mbox{opt}}}{\hbar \omega}\nonumber \\   &-&
 {1\over \hat \rho ^2} +
8\sqrt 2 \pi n\left|\frac {\varphi({\hat \rho,\hat y};t)}{\hat
\rho}\right|^2
 \biggr]\varphi({ \hat \rho,\hat y};t)=0,
\end{eqnarray}
where
$ n =   N a /l$. In terms of 
probability
 $P( y,t) \equiv$ $ 2\pi$ $\int_0 ^\infty
d\hat \rho |\varphi(\hat \rho,\hat y,t)|^2/\hat \rho $, the normalization
of the wave function
is $\int_{-\infty}^\infty d\hat y P(y,t) = 1.$ 
The probability
$P(y,t)$ is  useful in the study
 of the formation and
evolution of the interference pattern and  Josephson
oscillation.

In the  experiments of Cataliotti {\it et al.} \cite{2,3}
with repulsive $^{87}$Rb atoms, the axial and radial trap frequencies were
$\nu \omega =
2\pi \times 9 $ Hz and $ \omega =
2\pi \times 92$ Hz, respectively. The
optical
potential created with the standing-wave laser field of wave length
$\lambda=795$ nm is given by $V_{\mbox{opt}}=V_0E_R\cos^2 (\kappa_Lz)$,
with $E_R=\hbar^2\kappa_L^2/(2m)$, $\kappa_L=2\pi/\lambda$ and $V_0$ $
(<12)$
the
strength. 
In terms of the dimensionless laser wave
length $\lambda _0= \sqrt2\lambda/l \simeq 1$, $E_R/(\hbar \omega)= 4\pi^2/\lambda _0^2$.
Hence  $V_{\mbox{opt}}$ of
Eq.  (\ref{d1}) becomes 
\begin{equation}\label{pot}
\frac{ V_{\mbox{opt}}}{\hbar \omega}=V_0\frac{4\pi^2}{\lambda_0^2}
\left[
\cos^2 \left(
\frac{2\pi}{c\lambda_0}y
\right)
 \right],
\end{equation}
where the parameter $c$  controls the
spacing
$c\lambda_0/2$
between the optical-lattice sites.  The experimental condition  of
Cataliotti {\it et al.} \cite{2,3} is obtained by taking $c=1$.


The GP equation (\ref{d1}) is solved by the Crank-Nicolson method
\cite{11}. 
An interference pattern is formed by suddenly removing the combined traps
at time $t=0$ on the ground-state solution.  To study the time evolution
of the
system  we plot in Figs. 1 (a), (b), (c) and  (d)  $P(y,t)$ vs.
$y$ and $t$ for $c=1, 2.4,3.5$ and  4.6, respectively. The variation
of $c$ of Eq. (\ref{pot}) corresponds to a
variation
of
the spacing between successive sites. The increase in $c$ simulates an
increase in the distance between the  sites and
a decrease in the number of occupied sites.  In these
plots one can clearly see the central condensate and the moving
interference peak(s).  As the number of occupied optical-lattice sites
decreases with the increase of $c$, the interference pattern develops more
and more peaks.        

\begin{figure}
\begin{center}
\includegraphics[width=.45\linewidth]{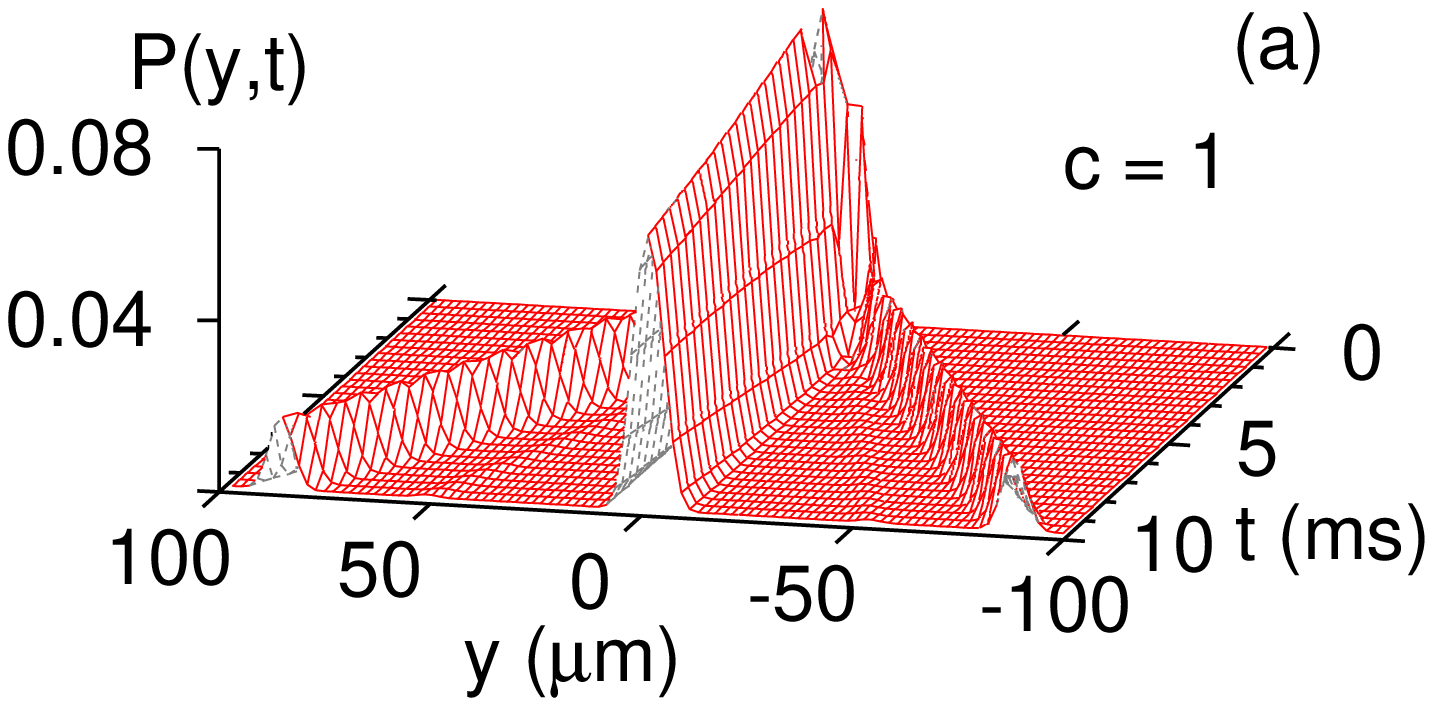}
\includegraphics[width=.45\linewidth]{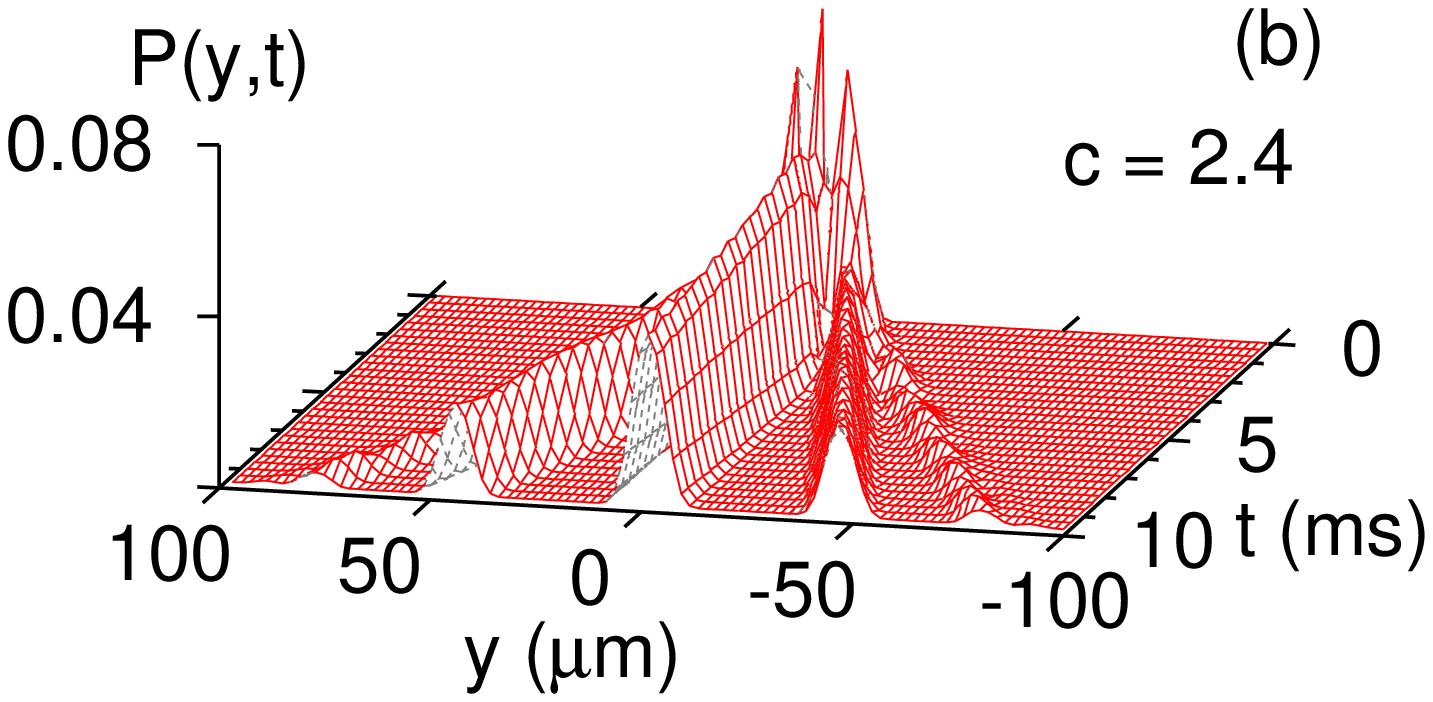}
\includegraphics[width=.45\linewidth]{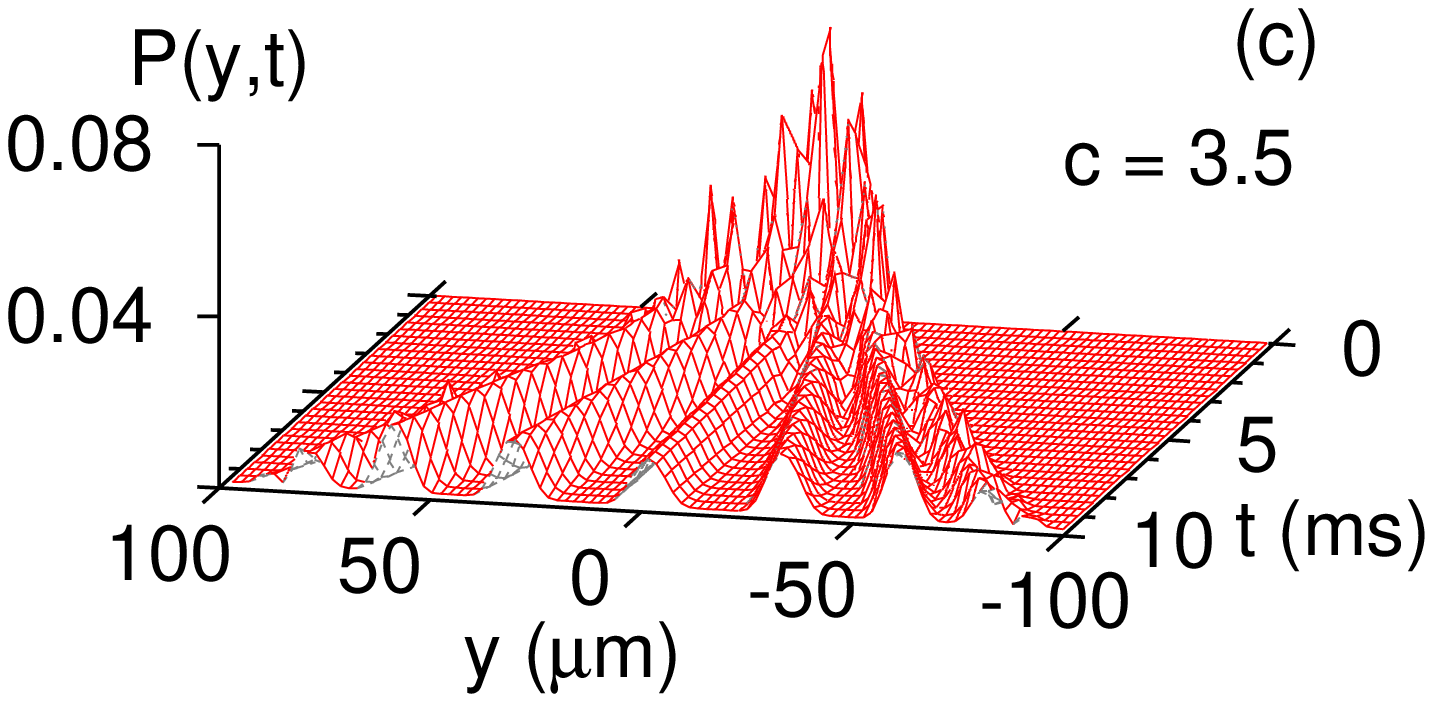}
\includegraphics[width=.45\linewidth]{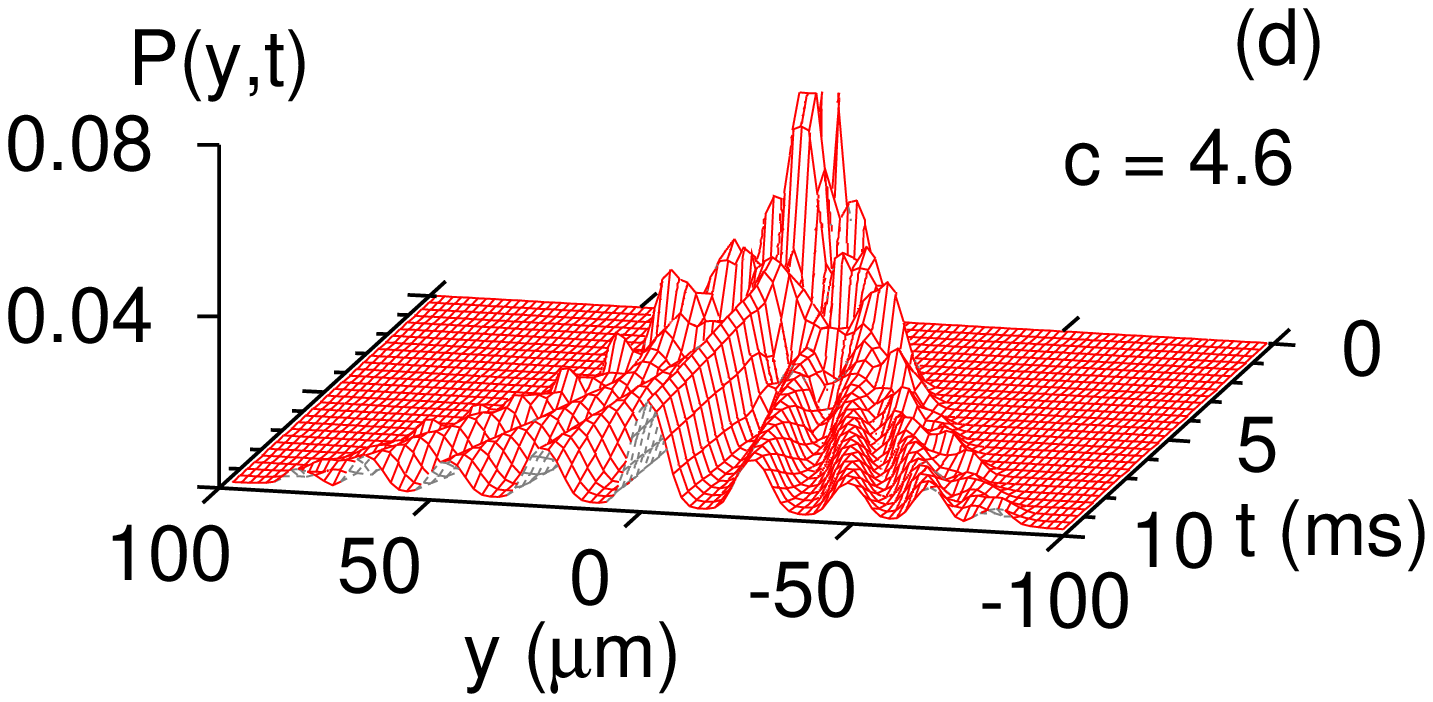}
\end{center}
\caption{$P(y,t)$
vs. $y$ and $t$ for the trapped   BEC
 after the removal of
combined optical and harmonic
 traps at $t=0$ for
  (a) $ c=1$,
(b) $c=2.4$, (c) $c=3.5$ and (d) $c=4.6$. The
interference
pattern has led to 3, 5, 7 and 9  peaks in these cases.}
\end{figure}

\begin{figure}[htb]
\begin{minipage}[t]{78mm}
\includegraphics[width=1\linewidth]{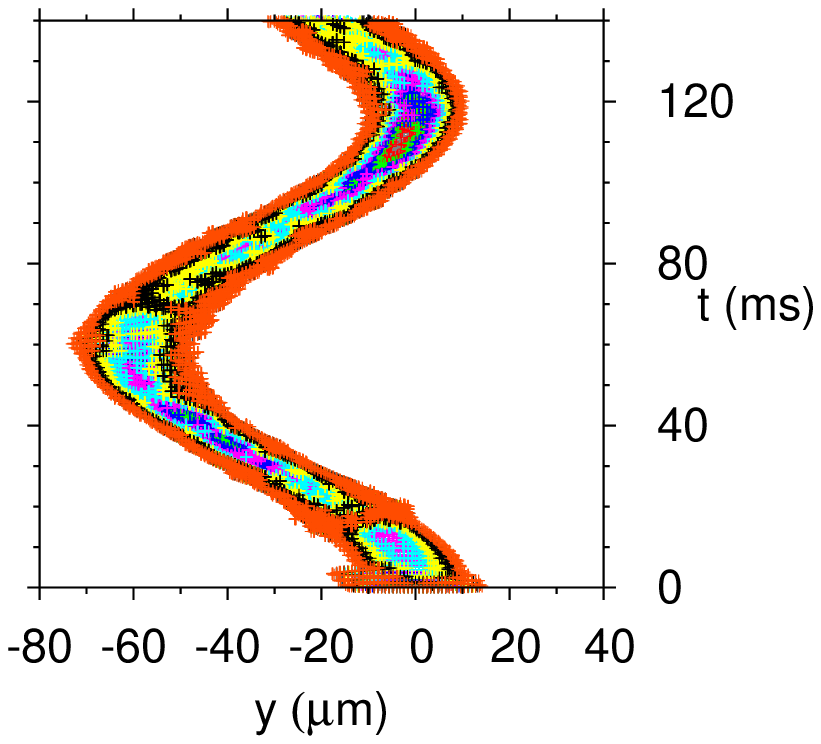}     
\caption{Contour plot of $P(y,t)$ vs. $y$ and $t$ for $V_0=2 E_R$ after a
harmonic
trap displacement of 30 $\mu$m 
showing the Josephson oscillation with frequency 8.45 Hz.}
\end{minipage}
\hspace{\fill}
\begin{minipage}[t]{75mm}
\includegraphics[width=.95\linewidth]{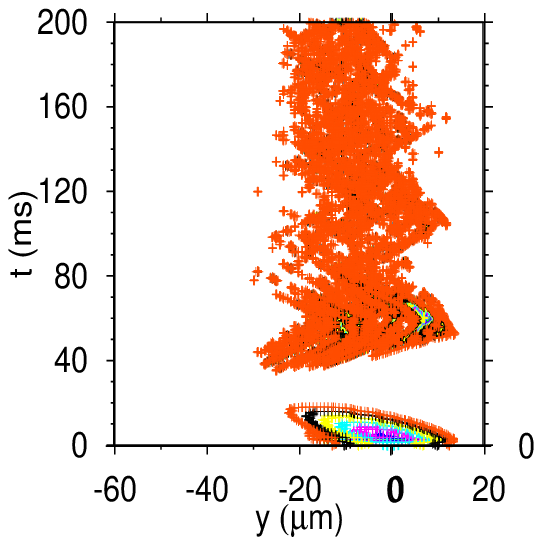}     
\caption{Contour plot of $P(y,t)$ vs. $y$ and $t$ for $V_0=5E_R$ after a
harmonic
trap displacement of 70 $\mu$m
demonstrating the disruption of the Josephson oscillation.}
\end{minipage}
\end{figure}
\begin{figure}[htb]
\begin{minipage}[t]{78mm}
\includegraphics[width=.9\linewidth]{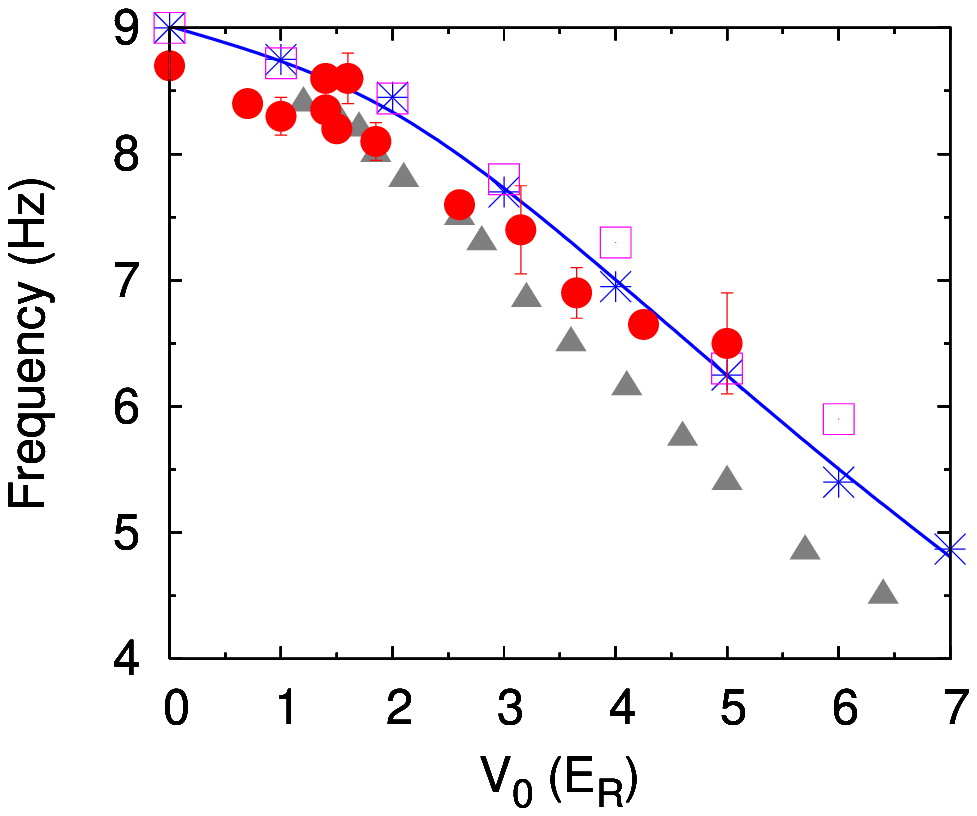}     
\caption{Josephson frequency vs. $V_0$: solid circle  with error bar -
experiment \cite{2}; solid triangle - tight-binding approximation
\cite{2}; square -
indirect result after expansion \cite{7}; star with full line - present
direct
result.}
\end{minipage}
\hspace{\fill}
\begin{minipage}[t]{78mm}
\includegraphics[width=.85\linewidth]{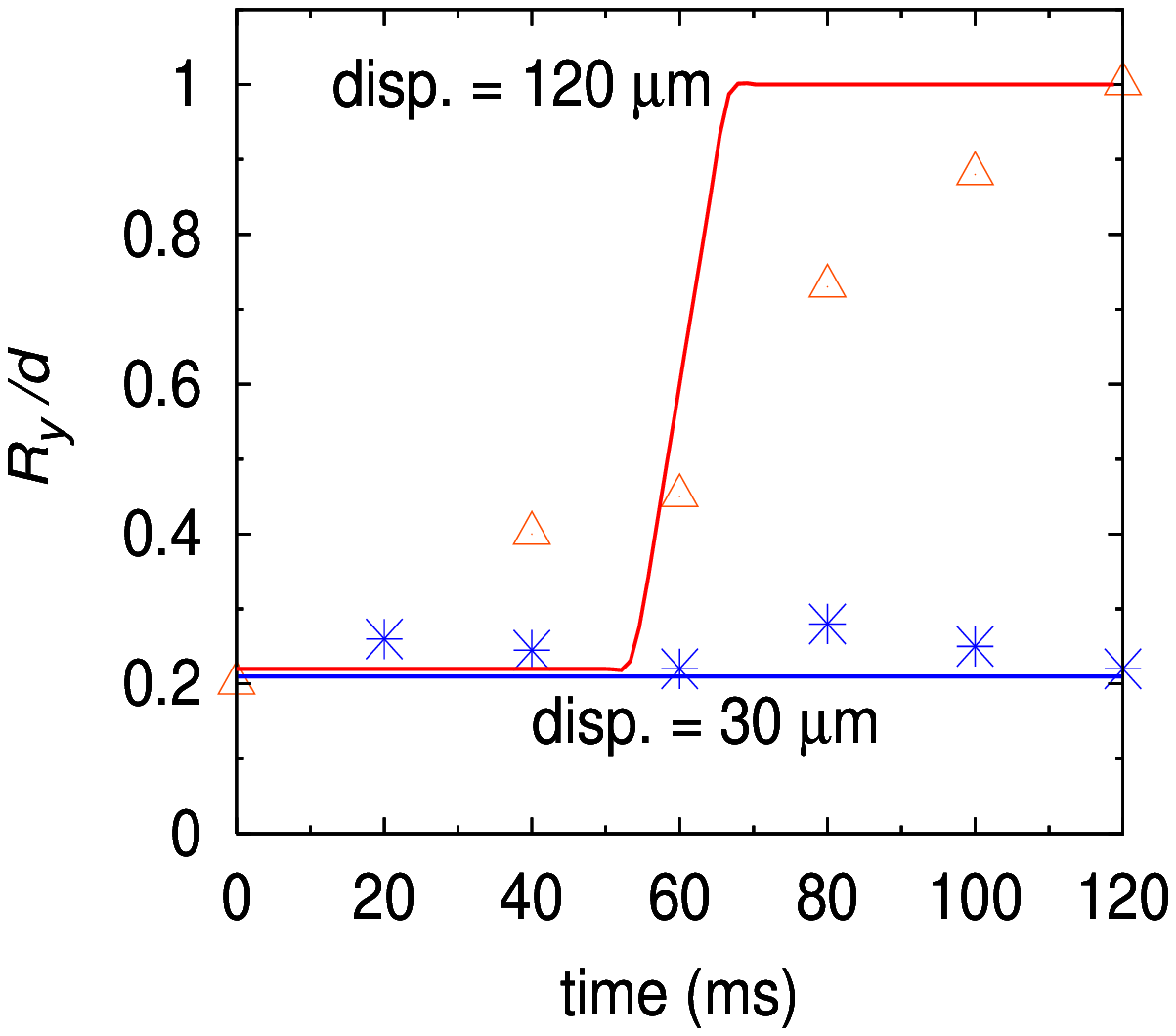}     
\caption{Axial width ($R_y$) of central  peak by peak 
separation $d$
vs. time spent in  displaced
trap: experiment  \cite{3} $-$  star -  displacement of 30 $\mu$m, 
triangle -
displacement of 120  $\mu$m;  present
theory $-$ full lines.} \label{fig:toosmall}
\end{minipage}
\end{figure}

We have illustrated in Fig. 1 the formation of interference pattern upon
releasing the BEC from the joint optical and harmonic traps. The formation
of interference pattern implies phase coherence and superfluidity in the
condensate. The Josephson oscillation is a direct manifestation of
superfluidity while the condensed atoms freely tunnel through the high
optical lattice barriers. The absence of  interference pattern after
displacing the harmonic trap implies the loss of superfluidity and a
disruption of the Josephson oscillation. This phenomenon represents a
superfluid to a classical insulator transition and can be studied by a
mean-field approach. The superfluid to a quantum Mott insulator transition
as in Ref. \cite{5} can only be understood by a field-theoretic approach
beyond mean-field theory.
We study the Josephson oscillation using both approaches, e.g.,
upon releasing the BEC from the joint traps as in the previous study
\cite{7} 
and by
following the condensate directly after displacing the harmonic
trap. Although a free expansion is the only way to observe the Josephson
oscillation experimentally, for numerical purpose the direct approach
seems to be more precise and involves less computer memory and time.

        In Fig. 2 we exhibit a contour plot of $P(y,t)$ vs. $y$ and $t$
for $V_0=2E_R$ after displacing the harmonic trap by a distance 30 $\mu$m
as in the experiment \cite{2}. The Josephson sinusoidal oscillation 
around the displaced trap center at $y=-30$ $\mu$m  is clearly
visible in this plot from which  the frequency of oscillation can be
obtained reasonably accurately. However, when the 
displacement of the
harmonic trap is increased beyond a critical value, 
the oscillatory motion
is disrupted as shown in Fig. 3 for   $V_0=5E_R$  and a  harmonic trap
displacement 
of 70
$\mu$m in agreement with experiment \cite{3}. In this case, unlike in
Fig. 2,  
the condensate does not cross the center of the
displaced trap at  $y=-70$
$\mu$m.
We performed a direct study of the 
Josephson 
oscillation for different  $V_0$ for a displacement of the harmonic trap
below the critical value for the disruption of superfluidity. In Fig. 4
we plot the Josephson frequencies calculated from this study 
as well as those calculated by allowing an expansion of the BEC as
in Ref. \cite{7}. In this figure we also plot \cite{2} the experimental
results as
well
as those obtained by using the tight-binding approximation. The
three-dimensional results obtained after expansion in Ref. \cite{7} and
obtained
directly here are in agreement with each other as well as with
experiment  \cite{2}. The present direct results
fit a smooth line and hence seem to be more accurate than the
results of  Ref. \cite{7}. The results for
tight-binding approximation \cite{2} are slightly different from the full 
three-dimensional results.

Finally, in Fig. 5 we plot the axial width $R_y$ of the central peak
normalized
to peak separation $d$  for $V_0=5E_R$      
vs. time spent in the displaced trap for displacements of 30 $\mu$m and
120 $\mu$m and compare with 
experiment \cite{3}. For the displacement of 30  $\mu$m, $R_y/d$ remains
constant,
whereas, for 120 $\mu$m,  $R_y/d$ increases to unity with time. However,
the
theoretical increase is much faster than in experiment.

To conclude, we have provided an account of matter-wave interference,
Josephson oscillation and its disruption using the
three-dimensional mean-field GP
equation. The results are in agreement with recent
experiments by Cataliotti {\it et al.} \cite{2,3}. The present
results for Josephson frequencies are slightly different
from those of tight-binding approximation. Further studies in three
dimension are needed to understand this difference.

\end{document}